\documentclass[sigconf,nonacm]{acmart}


\usepackage[utf8]{inputenc}
\usepackage{hyperref}
\usepackage{listings}
\usepackage{booktabs}
\usepackage{multirow}
\usepackage{makecell}
\usepackage[htt]{hyphenat}
\usepackage{courier}
\usepackage{balance}

\usepackage{listings, xcolor}

\definecolor{verylightgray}{rgb}{.97,.97,.97}

\lstdefinelanguage{Solidity}{
	keywords=[1]{anonymous, assembly, assert, balance, break, call, callcode, case, catch, class, constant, continue, constructor, contract, debugger, default, delegatecall, delete, do, else, emit, event, experimental, export, external, false, finally, for, function, gas, if, implements, import, in, indexed, instanceof, interface, internal, is, length, library, log0, log1, log2, log3, log4, memory, modifier, new, payable, pragma, private, protected, public, pure, push, require, return, returns, revert, selfdestruct, send, solidity, storage, struct, suicide, super, switch, then, this, throw, transfer, true, try, typeof, using, value, view, while, with, addmod, ecrecover, keccak256, mulmod, ripemd160, sha256, sha3}, 
	keywordstyle=[1]\color{blue}\bfseries,
	keywords=[2]{address, bool, byte, bytes, bytes1, bytes2, bytes3, bytes4, bytes5, bytes6, bytes7, bytes8, bytes9, bytes10, bytes11, bytes12, bytes13, bytes14, bytes15, bytes16, bytes17, bytes18, bytes19, bytes20, bytes21, bytes22, bytes23, bytes24, bytes25, bytes26, bytes27, bytes28, bytes29, bytes30, bytes31, bytes32, enum, int, int8, int16, int24, int32, int40, int48, int56, int64, int72, int80, int88, int96, int104, int112, int120, int128, int136, int144, int152, int160, int168, int176, int184, int192, int200, int208, int216, int224, int232, int240, int248, int256, mapping, string, uint, uint8, uint16, uint24, uint32, uint40, uint48, uint56, uint64, uint72, uint80, uint88, uint96, uint104, uint112, uint120, uint128, uint136, uint144, uint152, uint160, uint168, uint176, uint184, uint192, uint200, uint208, uint216, uint224, uint232, uint240, uint248, uint256, var, void, ether, finney, szabo, wei, days, hours, minutes, seconds, weeks, years},	
	keywordstyle=[2]\color{teal}\bfseries,
	keywords=[3]{block, blockhash, coinbase, difficulty, gaslimit, number, timestamp, msg, data, gas, sender, sig, value, now, tx, gasprice, origin},	
	keywordstyle=[3]\color{violet}\bfseries,
	identifierstyle=\color{black},
	sensitive=false,
	comment=[l]{//},
	morecomment=[s]{/*}{*/},
	commentstyle=\color{gray}\ttfamily,
	stringstyle=\color{red}\ttfamily,
	morestring=[b]',
	morestring=[b]"
}

\title[Using SPLs to Create Blockchain Products: Application to Supply Chain Traceability]{Using Software Product Lines to Create Blockchain Products: Application to Supply Chain Traceability}

\author{Nicolas Six}
\email{nicolas.six@univ-paris1.fr}
\orcid{0000-0001-7563-3628}
\affiliation{%
  \institution{Université Paris 1 Panthéon-Sorbonne}
  \streetaddress{90, Rue de Tolbiac}
  \city{Paris}
  \country{France}
  \postcode{75013}
}

\author{Nicolas Herbaut}
\email{nicolas.herbaut@univ-paris1.fr}
\affiliation{%
  \institution{Université Paris 1 Panthéon-Sorbonne}
  \streetaddress{90, Rue de Tolbiac}
  \city{Paris}
  \country{France}
  \postcode{75013}
}

\author{Roberto Erick Lopez-Herrejon}
\email{roberto.lopez@etsmtl.ca}
\affiliation{%
  \institution{École de Technologie Supérieure}
  \streetaddress{1100 Notre-Dame St W}
  \city{Montreal}
  \country{Canada}
  \postcode{H3C 1K3}
}

\author{Camille Salinesi}
\email{camille.salinesi@univ-paris1.fr}
\affiliation{%
  \institution{Université Paris 1 Panthéon-Sorbonne}
  \streetaddress{90, Rue de Tolbiac}
  \city{Paris}
  \country{France}
  \postcode{75013}
}

\date{March 2022}

\begin{document}

\begin{abstract}
  In recent years, blockchain has been growing rapidly from a niche technology to a promising solution for many sectors, due to its unique properties that empower the design of innovative applications. 
  Nevertheless, the development of blockchain applications is still a challenge. 
  Due to the technological novelty, only a few developers are familiar with blockchain technologies and smart contracts. 
  Others might face a steep learning curve or difficulties to reuse existing code to build blockchain applications. 
  This study proposes a novel approach to tackle these issues, through software product line engineering. 
  To support the approach, a web platform to configure and generate a blockchain application for on-chain traceability is introduced.
  First, a feature model has been designed to model core features of the chosen domain, based on the existing literature. 
  Then, a configurator has been implemented to support the feature selection phase. 
  Finally, a generator is able to ingest such configurations to generate on-the-shelf blockchain products. 
  The generalizability of the contribution is validated by reproducing on-chain traceability applications proposed in the literature by using the platform. 
  This work provides the first evidence that the implementation of blockchain applications using software product lines enhances the quality of produced applications and reduces the time to market.
\end{abstract}


\maketitle

\section{Introduction}

While blockchain's popularity growth coupled with its unique capabilities has attracted many companies to start blockchain projects, the road going from their drive to innovate to the materialization into production-ready applications remains challenging. 
Three issues hinder the adoption of blockchain: organizational, legal, and technological \cite{prewett2020blockchain}. 
For the latter, one reason is the difficulties met by software developers along the software engineering process. 
The design of the application can be tedious due to the novelty of components employed inside, such as smart contracts, or cryptographic wallets. 
A bad design can lead to higher operation and maintenance costs at best and vulnerabilities and flaws at worst (e.g. \textit{The DAO} attack \cite{mehar2019understanding}).

Reusing existing code is one solution to solve this issue (so-called clone-and-own) and is a common practice in the blockchain field \cite{chen2021understanding}.
Some of these solutions have even been formalized as design patterns to ease their reuse.
For instance, as smart contracts cannot query data from outside the blockchain, developers have to apply a design pattern named \textit{Oracle pattern} \cite{xu2018pattern}. 
An oracle includes two components: a smart contract capable of emitting an event when new data is required, and an off-chain service listening to these events to inject fresh data when needed.

This reuse of existing code is a first step in addressing the difficulties of implementing a blockchain application, but it could be further systematized with a software product line (SPL) approach.
Software product line engineering (SPLE) is based on the reuse of various software artifacts (e.g., requirements, models, code, and tests) designed for this purpose, to create (software) products that have common elements \cite{pohl2005software}.
By leveraging an SPL approach, developers could easily configure and generate blockchain applications by reusing existing knowledge and artifacts.
First, by reusing domain requirements to guide the selection of features in a feature model (introduced in Section \ref{feature-model}). 
Then, by reusing design artifacts, such as existing design patterns or models.
Finally, by reusing configurable code (in this case, templates) that may also implement the aforementioned design patterns.
Instead of using the "clone-and-own" strategy, configurable code can be tailored to fit into the project systematically.
Yet, the combination of SPL and blockchain technology is still an unexplored area.
This study attempts to highlight the relevance of coupling SPLs and blockchain by addressing the following research questions:

\begin{itemize}
  \label{research-questions}
  \item \textbf{RQ1} - Is SPLE applicable to the blockchain field?
  \item \textbf{RQ2} - Do blockchain applications created following a standard software development engineering differs from applications derived from a SPL?
\end{itemize}

To answer these questions, this contribution proposes the creation from scratch of an SPL for blockchain applications.
It results in a web platform that allows the configuration and the generation of a blockchain product.
The generation is performed by assembling code templates (e.g., smart contracts), based on the configuration given by the user.
A feature model guides the configuration process, by describing existing features and their constraints with others.
This feature model has been devised by extracting features found in studies of a specific domain, that is, blockchain-based traceability. 
We evaluated the capacity to generalize our approach by reproducing existing blockchain-based traceability applications using exclusively the web platform.
Also, the source code of the web platform and the templates is available on Github\footnote{\url{https://github.com/harmonica-project/BANCO}}

The paper is organized as follows: Section \ref{background} introduces background on blockchain technologies, then Section \ref{related-works} discusses related works on blockchain code generation and variability.
Section \ref{feature-model} and \ref{platform-construction} introduce the platform, with describing the construction of the feature model and its reuse through the web platform.
An evaluation is performed in Section \ref{evaluation}, and Section \ref{discussion} discusses those results along with lessons learned and open research challenges.
Finally, Section \ref{conclusion} concludes the paper.

\section{Background \label{background}}

In this section, some background on blockchain technologies and smart contracts is given.
A blockchain is a data structure where each block is linked to the previous one with a cryptographic hash.
Each block also contains a list of transactions that represent user-to-blockchain interactions.
A network of peers, known as nodes, is in charge of adding new blocks to the blockchain.
First-generation blockchains, such as Bitcoin \cite{nakamoto2008bitcoin}, only had the purpose of cryptocurrency exchange between users.
Nevertheless, the release of Ethereum \cite{wood2014ethereum} in 2015 allowed its usage in a wider range of use cases through Turing-complete smart contracts.

A smart contract (also called DApp) is a computer program that executes predefined actions when certain conditions within the system are met.
Blockchain smart contracts can be deployed and interacted with through transactions.
The nodes responsible for the inclusion of new transactions into the blockchain are responsible for executing the smart contract with provided parameters.
Each smart contract is constituted of two distinct parts: its state, and its logic.
By interacting with smart contract functions using transactions, it is possible to alter the state of a smart contract.

The usage of smart contracts combined with blockchain allows the development of applications that differ from conventional software engineering.
These applications benefit from blockchain decentralization, as there is no central actor to control the network.
Smart contract data are also immutable by nature, as it is theoretically impossible to alter a block after its addition into the network.
Finally, full transparency of the application is possible, allowing the traceability of its data and usages. 
Many usages of blockchain have already been explored in the literature.
For instance, Tian et al. propose to use blockchain for transparent and trustable traceability of supply chain \cite{tian2016agri}.
Decentralized finance (DeFi) leverages blockchain to exchange value between users without any intermediary \cite{schar2021decentralized}.
Blockchain can also be used as a platform for peer-to-peer systems, such as smart grids \cite{mollah2020blockchain}. 

\section{Related Work \label{related-works}}

Several works in the literature have been proposed to assist practitioners in designing, generating, and deploying blockchain-based solutions, starting from low-level code generation tools to Model-driven Engineering (MDE) and proposals that take blockchain variability into account.

\subsection{Smart Contract Code Generation}
 The most recent blockchain solution supports general-purpose programming languages, such as Java/Kotlin for Corda \cite{hearn2016corda}, or Go/Node.js/Java for Hyperledger Fabric \cite{androulaki2018hyperledger}. 
 Yet, the vast majority of the literature presenting blockchain-based solutions still rely on Ethereum and its specific languages (e.g. Solidity) to demonstrate the feasibility of their proposal.
 For this reason, several papers focus on helping developers write smart contracts with Ethereum.

 Wöhrer et al. propose a Contract Modeling Language (CML) to simplify the writing of smart contracts \cite{Wohrer_2020}.
 CML defines contract-specific concepts such as Party, Asset, or Event, and decorators to indicate the usage of blockchain-based design patterns in specific functions.
 A parser is also proposed for CML-to-Solidity conversion.
 However, this approach requires developers to become proficient in CML in addition to Solidity, as only learning CML might limit developers in the development of smart contracts.
 Other approaches in the literature focus on reusing existing models to generate code.
 For instance, Zupan et al. propose a framework to generate smart contracts based on Petri nets \cite{zupan2020secure}.
 The generation of code is made through their translation engine, which is able to convert Petri nets into Solidity smart contracts. 
 López-Pintado et al. use Business Process Model and Notation (BPMN) to generate a suite of Solidity smart contracts, that are able to run the corresponding business process on the blockchain with a solution called Caterpilar \cite{lopez2019caterpillar}.
 Generated smart contracts are used to start business process instances, manage business process activities, and handle the business process workflow. 
 Choudhury et al. use a different model for smart contract generation composed of an ontology with classes linked together, and constraints expressed as a set of rules \cite{choudhury2018auto}. 

 \subsection{Blockchain and Model-Driven Engineering}

Smart contract code generation is useful for use cases where all the processed data happens to be on the blockchain. However, these approaches fall short when dealing with the integration of other domain-specific components into the blockchain solution at different architectural levels.
Several authors propose relying on MDE to help grasp the complexity of integrating blockchain-based solutions within information systems.

Lu et al. propose a tool called Lorikeet that extends the BPMN modeling capabilities already proposed in Caterpillar with the support of asset registry management~\cite{Lu_2020}.
Both business process modeling and asset registry modeling are used to generate smart contracts making the developers more productive, the operators able to monitor smart contracts execution, and the domain experts capable of understanding how their ideas are represented in the system.
De Sousa et al. present MDE4BBIS, a framework to incorporate MDE in the development of Blockchain-based IS \cite{de_Sousa_2021}.
They demonstrate their solution to support cross-organizational business processes.
Górski et al. propose new UML stereotypes in a UML profile for distributed ledger deployment and incorporated their solution in a modeling tool to automate the deployment to Corda \cite{Gorski_2020}.

\subsection{Blockchain and SPL}

Finally, a few proposals have been made to use SPLs for blockchain.
Kim et al. present a feature model to allow organizations to build their blockchain platform by selecting its features (e.g., smart contract language, consensus algorithm, etc.) \cite{Kim_2018}.
They present a feature model for blockchain platforms allowing the selection of the desired features, without, however, supporting feature binding or code generation.
Liaskos et al. introduce a meta-model for the derivation of specialized blockchain network simulators, emphasizing the importance of SPLE and MDE \cite{Liaskos_2020}.
As we have seen from this section, even if code generation and MDE have been proposed to support the creation of blockchain applications, this paper is the first attempt at building an SPL for blockchain applications.

\section{Feature Model Design}
\label{feature-model}

The first step in the SPLE process is the domain analysis \cite{czarnecki2000generative}, where the result is often a feature model.
A feature model (FM) is a widely adopted notation to describe allowed variability between products of the same family and feature dependencies \cite{schobbens2007generic}.
The main advantage of using a FM is the increased ease of reusing existing features, with an accurate mapping that can be shared between stakeholders.
In this study, the FM has been created following the standard feature model notation with FeatureIDE, an open-source framework \cite{thum2014featureide}.
It is composed of different notation elements.
It allows the definition of concrete/abstract features that can be optional or mandatory.
It also supports \textit{and-} and \textit{xor-} decomposition of features, to either select multiple subfeatures (but at least one) among a given set linked to a feature, or select only one subfeature in the selection.
Finally, FMs include constraints between features, preventing for instance the selection of two conflicting features.
The standard FM has been chosen as it satisfies our needs for the construction of an on-chain traceability FM.

\subsection{Construction Method}
%
\begin{table*}[t]
  \caption{Blockchain traceability research used to design and test the FM}
  \label{tab:fms}
  \begin{tabular}{|l|p{10.8cm}|l|c|c|c|}
  \hline
  \textbf{Ref} & \textbf{Title} & \textbf{Authors} & \textbf{Item} & \multicolumn{1}{c|}{\textbf{Part of}} \\ \hline
  \cite{baralla2021ensuring} & Ensuring transparency and traceability of food local products: A blockchain application to a Smart Tourism Region. & Baralla et al. & Food & \multirow{5}{*}{\makecell{Founda-\\tional\\set}} \\ \cline{1-4}
  \cite{caro2018blockchain} & Blockchain-based traceability in Agri-Food supply chain management: A practical implementation. & Caro et al. & Food & \\ \cline{1-4}
  \cite{figorilli2018blockchain} & A blockchain implementation prototype for the electronic open source traceability of wood along the whole supply chain. & Figorilli et al. & Wood  &  \\ \cline{1-4}
  \cite{kuhn2021blockchain} & Blockchain-based application for the traceability of complex assembly structures & Kuhn et al. & \makecell{Manufactured\\items} & \\ \cline{1-4}
  \cite{wei2020blockchain} & Blockchain-based data traceability platform architecture for supply chain management & Wei et al. & Goods & \\ \hline
  \cite{hasan2020blockchain} & Blockchain-based solution for the traceability of spare parts in manufacturing & Hasan et al. & \makecell{Spare parts} & \multirow{2}{*}{Test set} \\ \cline{1-4}
  \cite{casino2021blockchain} & Blockchain-based food supply chain traceability: a case study in the dairy sector & Casino et al. & Food & \\ \hline
  \end{tabular}
  \end{table*}

The construction of a FM requires extensive knowledge of its associated domain.
In this study, this knowledge has been extracted from 5 different works that propose on-chain traceability solutions, called foundational set, shown in Table~\ref{tab:fms} \cite{baralla2021ensuring, wei2020blockchain, caro2018blockchain, figorilli2018blockchain, kuhn2021blockchain}.
From these papers, the features that were at least present twice (2-of-5) were included in the FM. 
In some cases, they have been refined manually by adding subfeatures (e.g., adding CRUD methods to manage application participants).
This results in a FM, presented in the following subsections.
Note that this FM is not meant to be a complete representation of existing on-chain traceability features, but provides its most salient characteristics.
A complete analysis through a systematic literature review is left for future work.

The resulting FM is composed of 53 different features, split across three different core features.
The first core feature, \texttt{SmartContracts}, gather features included in the smart contracts.
The selection of the subfeatures of \texttt{SmartContracts} represents the configuration of the on-chain part of the application.
The second core feature, named \texttt{Storage}, regroups the features that address how and where traceability data is stored.
Finally, the last core feature, \texttt{Frontend}, represents the off-chain part of the application.

\subsection{Feature Smart Contracts}

\begin{figure}[h]
  \centering
  \includegraphics[width=\linewidth]{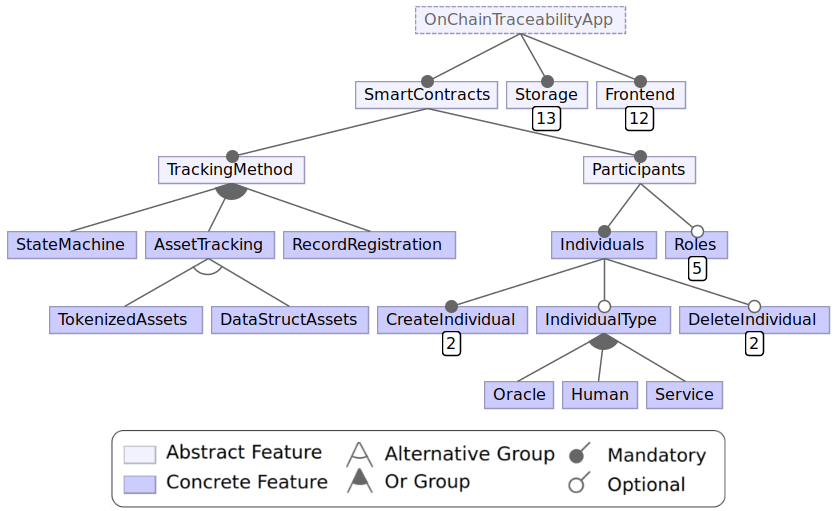}
  \caption{Focused view of the SmartContract FM}
  \label{fm-sc}
\end{figure}

The first feature of the model is \texttt{SmartContracts} (Figure \ref{fm-sc}).
It represents the on-chain part of the traceability application, composed of a collection of smart contract instances.
This part of the FM also involves three different constraints, expressed in Table \ref{c-fm}.

This feature is divided into two subfeatures: the management of participants, and the traceability methods used.
The feature \texttt{Participants} distinguish two important aspects: individuals, that will interact with the traceability smart contracts and are identified by a public address, and roles, that can be assigned to individuals.
Using roles is optional in the model, as access control can be done using only public addresses (e.g., only a given set of individuals can add records in a given record collection).
However, they can be useful to implement Role-based Access Control (i.e., in a supply chain, identify the suppliers, carriers, and buyers).
Besides roles, individuals can be classified through types: they can either be human, service, or oracle (from the \textit{Oracle pattern} \cite{xu2018pattern}).

Three traceability methods can be selected in conjunction or as standalone in the model.
The subfeature \texttt{StateMachine} allows tracking state changes on-chain.
A state machine is defined by a set of state variables and commands, that transform its state \cite{schneider1990implementing}.
For each transition, it is possible to define a set of individuals and roles that are entitled to trigger the transition between two states.
The current implementation behind the subfeature \texttt{StateMachine} only allows the creation of basic state machines where each state only has at most one previous and one following state.
Nonetheless, this aspect will be improved in future works.
\texttt{AssetTracking} consists of storing data on real-world assets.
Each asset has a set of owners and a set of "entitled" individuals and roles that can modify it.
A state machine can be attached to an asset: for instance, a batch can be stored, shipped, or delivered.
Assets can either simply be stored as a simple data structure, or as tokens (as proposed in \cite{kuhn2021blockchain}).
Storing assets as tokens facilitates their transfer between individuals.
For instance, a batch can be sent from the supplier to the carrier.
Tokenization is a common blockchain-based design pattern \cite{xu2018pattern}, standardized for many blockchains such as the ERC721 standard for Ethereum\footnote{\url{https://eips.ethereum.org/EIPS/eip-721}}.
Finally, \texttt{RecordCollections} allows bulk storage of records in arrays. 
These records are stored as described in the feature \texttt{Storage}.
As with others, a collection has a set of "entitled" individuals and roles that can append new records.

\begin{table}[t]
  \caption{Feature model constraints}
  \label{c-fm}
  \begin{tabular}{ccl}
    \toprule
    Range & Operator & Target \\
    \midrule
    DeleteIndividualByRole & $\Rightarrow$ & Roles \\
    IndividualsSetup & $\iff$ & CreateIndividualAtSetup \\
    RolesSetup & $\iff$ & CreateRoleAtSetup \\
    RecordRegistration & $\iff$ & RecordHistory\\
    RecordHistory & $\iff$ & RecordsCollectionSetup\\
    AssetTracking & $\iff$ & AssetsData\\
    AssetsData & $\iff$ & AssetsSetup\\
    StateMachine & $\iff$ & StateMachineData\\
    StateMachineData & $\iff$ & StateMachineSetup\\
  \bottomrule
\end{tabular}
\end{table}

\subsection{Feature Storage}

\begin{figure}[h]
  \centering
  \includegraphics[width=\linewidth]{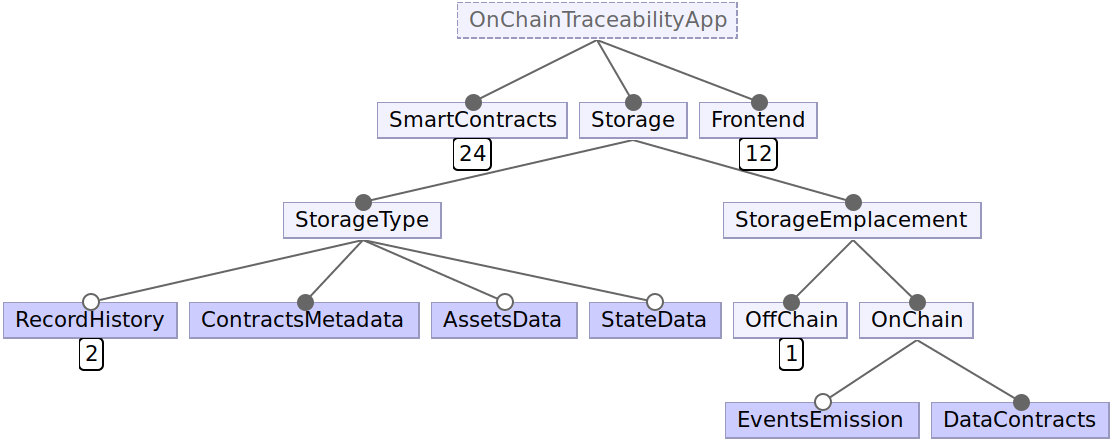}
  \caption{Focused view of the Storage FM.}
  \label{fm-storage}
\end{figure}

The second feature of this model is \texttt{Storage} (Figure \ref{fm-storage}), divided in two aspects.
For the first aspect, data can be stored in multiple formats.
In some applications, it is a suite of timestamped records.
These records can be either data on a specific event that occurred in the traceability process or regularly pushed traceability data (e.g., real-time temperature).

The FM further refines the subfeature \texttt{RecordHistory} in two: \texttt{StructuredRecords} and \texttt{HashedRecords} (not shown in Figure \ref{fm-storage}).
Where the first one can contain any type of data, the second one is a timestamped hash of a \texttt{StructuredRecord}.
These records can be used when it is not desirable to store data on-chain for confidentiality reasons or storage limitations.
In this case, each structured record is stored off-chain in a database, then hashed and stored on-chain as it.
This storage strategy is a common blockchain design pattern named \textit{Off-chain data storage Pattern} \cite{xu2018pattern}.
Traceability data can also be stored as objects representing \texttt{AssetsData}, or as a set of states and the transition history between them when using a \texttt{StateMachine}.
These dependencies between storage type and traceability methods imply a set of constraints (Table \ref{c-fm}).
Indeed, the selection of a specific traceability method should automatically select the related setup form and storage type features.
Finally, a mandatory feature named \texttt{ContractMetadata} is in charge of storing the address of every smart contract deployed for a traceability process.
This feature includes the usage of the \textit{Factory Pattern} \cite{xu2018pattern}, as the factory deploys and keeps track of existing contract instances.

Regarding the storage emplacement, data can either be stored on-chain or off-chain.
On-chain data is stored in smart contracts following the \textit{Data Contract Pattern}, that separates data storage from logic contracts (e.g., controllers) \cite{xu2018pattern}.
Events can also be emitted when something occurs (e.g., storing a new record, firing a transition).
Traceability data can also be stored off-chain, in databases.
The \texttt{Database} feature is mandatory in the FM, as smart contract metadata must at least be stored off-chain to allow retrieving the address of existing contracts.
However, traceability data can either be stored off-chain, on-chain, or both.

\subsection{Feature Frontend}

\begin{figure}[h]
  \centering
  \includegraphics[width=\linewidth]{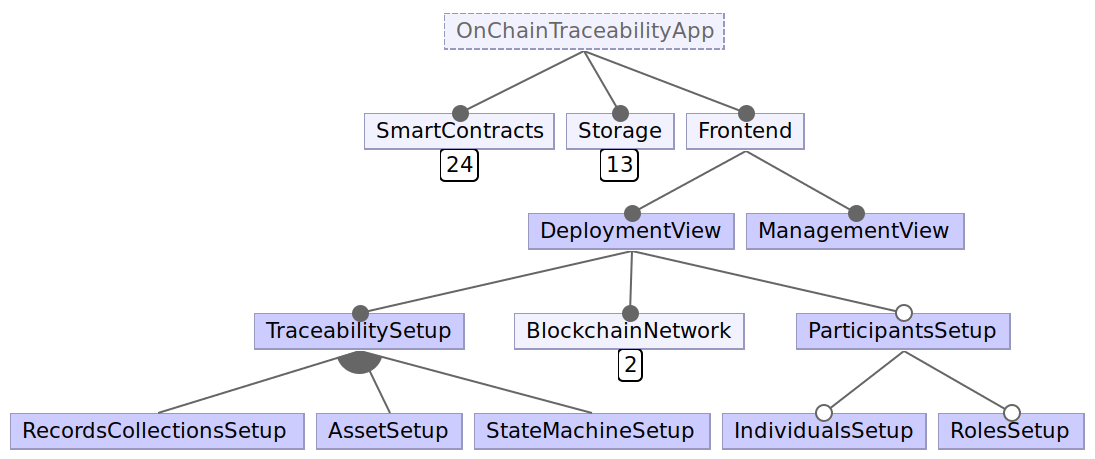}
  \caption{Frontend FM.}
  \label{fm-front}
\end{figure}

The last feature is \texttt{Frontend} (Figure \ref{fm-front}).
The frontend application can be used to set up the traceability process through the feature  \texttt{DeploymentView}.

Individuals, roles, and traceability assets/states/collections are not defined statically in the code but dynamically as parameters passed when instantiating the smart contracts.
Thus, the user has to specify these data to set up the traceability process.
One feature that is \texttt{BlockchainNetwork}, specifies the targeted network: in this model, either the Ethereum testnet (for testing purposes, free to use) or mainnet (in production).
Users can then interact with deployed smart contracts through the application to leverage the aforementioned features.

\section{Platform Construction}

\label{platform-construction}

The FM guides the possible selection of features by the user when configuring products.
However, this task is burdensome when performed manually.
In this work, a web platform has been built to ease the configuration of a product.
It notably integrates a configurator and a FM visualizer, that was adapted from Kuiter et al. work \cite{kuiter2018getting}.
This platform also integrates a generator based on templates to output a working product from the user configuration.
The following subsections respectively discuss the construction of the configurator and the generator.

\subsection{Product Configuration} 

We implemented two different panels on the web platform to ease the configuration.
The first panel displays a tree of features, generated using as input the on-chain traceability FM.
Some of the features are already pre-selected, as the FM contains mandatory features.
The user can either select the inclusion or the exclusion of a feature by selecting the corresponding box.
Each selection will trigger the constraint engine, which will automatically include or exclude features based on the constraints formulated along the FM.

The configuration also has two different states: its validity and its completeness.
The first one indicates if a configuration is valid, i.e., if constraints are satisfied.
As the configurator prevents selecting two conflictual features, the user cannot make a selection that results in an invalid configuration.
The second one indicates if the configuration is complete, i.e., all the features are either selected or deselected.
The user can rely on these indicators to know if the configuration step is complete or not.

The second panel shows the FM itself, to visualize the on-chain traceability domain and its available features. 
The visualizer guides the user during the configuration by changing the color of selected or deselected features respectively in green or red.
It allows to quickly notice the impact of selecting one feature on others, and the features that remain to be selected.

\subsection{Product Generation}

From a valid and complete product configuration, the web platform is capable of generating a working product leveraging a generator based on Template-Based Code Generation (TBCG).
TBCG is a technique from the MDE field that consists of generating code based on templates, constituted of static text with embedded dynamic portions that are evaluated by a template engine to output functioning code \cite{jorges2013construction}.
Such evaluation also requires providing data to fill the dynamic portions of the text.

In this work, the task of evaluating templates is performed by Mustache, a logic-less web template system\footnote{\url{https://mustache.github.io/}}.
Mustache is capable of evaluating any provided text input that contains a series of tags (i.e., dynamic portions), providing it a suitable JSON object to populate the tags with data.
This template system handles features such as optional code blocks, text completion, and loops.

From the configuration made by the user on the web platform, a JSON object is generated containing all of its choices, then ingested by Mustache to process the templates.
For the on-chain part, the smart contract templates are written in Solidity \cite{dannen2017introducing}, a language to implement Ethereum smart contracts.
The default Mustache tag has been modified from the default notation (\textit{\{\{ \}\}}) to the block comment symbols used in Solidity (\textit{/* */}), to allow writing Mustache instructions in Solidity comments.
This allows for developing and testing smart contract templates without raising any errors caused by the Mustache notation and preserves productivity-enhancing features of Integrated Development Environments such as static code analysis.
The approach taken to develop the templates is based on subtractive code generation: all of the features are included in the templates, and Mustache removes or modifies them according to the configuration.
For instance, the following code block (Listing \ref{code-sample}) will be conditionally rendered in the final product only if the feature \textit{AddRoleDynamically} has been selected by the user.

\begin{lstlisting}[caption={Solidity template code sample},label={code-sample},language=Solidity]
/* #AddRoleDynamically */
function addRoleToP(address _p, string _rName) public {
   [...]
}
/* /AddRoleDynamically */
\end{lstlisting}

\begin{figure}[t]
  \centering
  \includegraphics[scale=0.6]{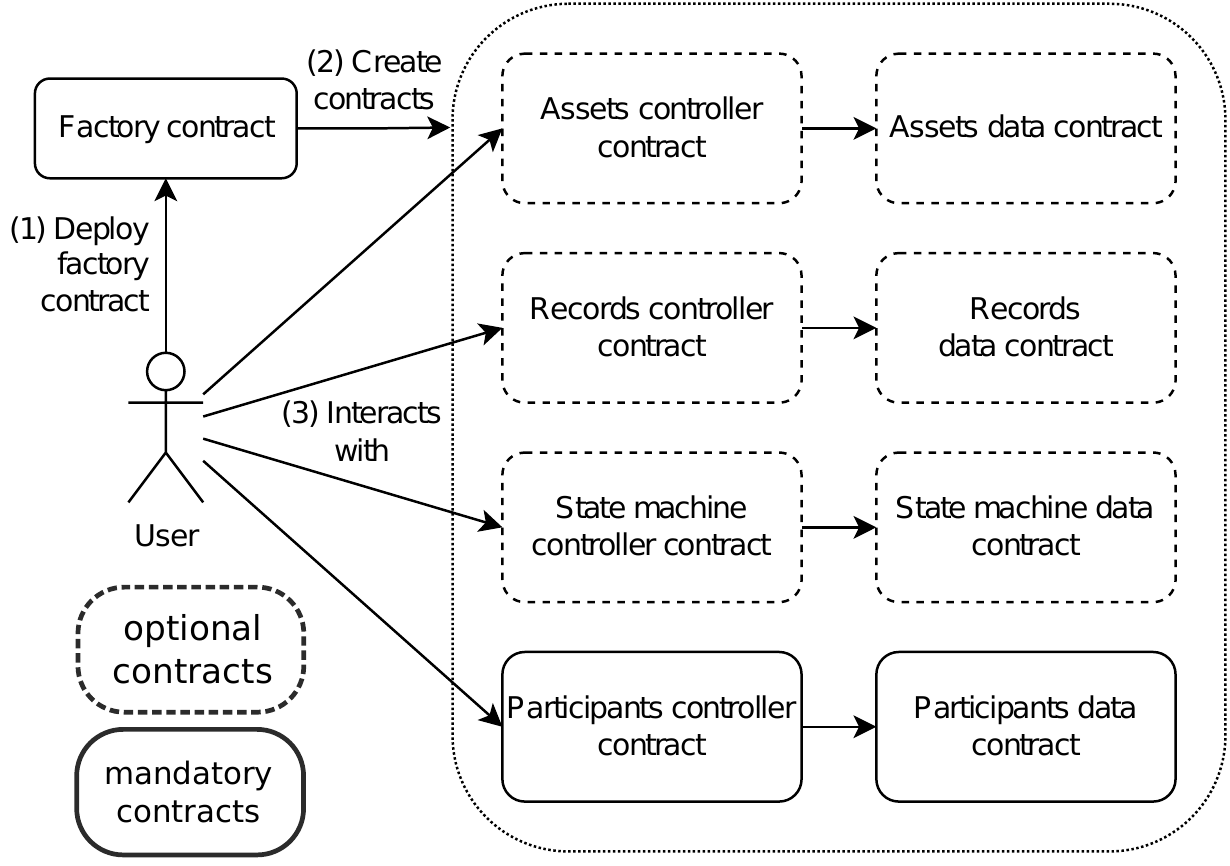}
  \caption{Smart contract architecture}
  \label{architecture}
\end{figure}

Figure \ref{architecture} describes the chosen architecture for the on-chain part of the application.
At first, the user deploys a single factory contract (1).
A factory contract, designed following the \textit{Factory pattern} \cite{xu2018pattern}, is in charge of creating other contract instances at instantiation (e.g., participant contracts) (2). 
The factory contract also acts as a contract registry, by storing created contracts' addresses.
Once this deployment is completed, the user can interact with controllers (3).
Each on-chain feature is implemented as a pair of two contracts: a data contract in charge of holding data collections and getters/setters to manipulate them, and a controller contract to interact with data contracts.
These controllers also enforce specific conditions to modify the data (e.g., verifying asset ownership before updating it).
The separation between logic and data is a common blockchain design pattern that also increases upgradeability: controllers can be changed without having to migrate the data from one contract to another \cite{xu2018pattern}.
Otherwise, this operation would be very expensive in terms of storage and costs and tedious to perform.

The different features defined in the FM can be traced to this architecture.
One controller/data smart contract pair is in charge of participants and roles, whereas three controllers/data contract smart pairs are responsible for the different traceability methods defined in the FM.
As the user can select one to three different traceability methods, some of these contracts might not be present in the final product.
However, the participant data/controller smart contract pair will always be present, although the role features might not be depending on the configuration.

For the off-chain part, a web application has been developed, where pages are conditionally included in the final product depending on the configuration.
For instance, if the user does not select the \texttt{Roles} feature, the web page to configure or allocate roles to users will not be included in the final product.

\section{Evaluation}
\label{evaluation}

The main motivations for using SPLs are the reduction of development costs, the reduction of time needed to create an application and an increased code quality \cite{pohl2005software}.
However, the produced artifacts using the SPL still have to satisfy the application requirements.
In this section, the requirements satisfaction and the cost of generated products are evaluated by being compared to reference implementations from a set of existing studies.

\subsection{Protocol}

The protocol used in the evaluation of the contribution is the following.
First, a sample of two studies has been chosen, called the Test Set in Table~\ref{tab:fms}.
This selection has been performed following two criteria: (1) the source code is available online and (2) the functional requirements of the application proposed in the study can be clearly identified and extracted.
Then, for each study, the following steps have been conducted:

\begin{enumerate}
  \item Extract main functional requirements\footnote{For the sake of brevity, we extracted only a subset of functional requirements that involve writing/modifying data.} formulated by the authors for their on-chain traceability application.
  \item Configure and generate a product using the web platform from these requirements.
  \item Assess the satisfaction of formulated requirements towards the produced blockchain application.
  \item Compare the operating cost of deploying the on-chain part of the product and the implementation proposed by the authors.
\end{enumerate}

During the second step, the configuration of a product is guided by the functional requirements.
However, some features are left to be configured at the end, as these features do not have any impact on the satisfaction of the requirements.
To arbitrate on these features, the source code of the reference paper implementation has been used to extend these requirements.
Finally, if some features remain unselected after the configuration, they are automatically deselected.
Indeed, in the Ethereum ecosystem, implementing additional features increases the operational cost of smart contracts.
Only one product is generated for each reference paper, however, many more products that also suit the specified requirements could have been generated.
Nonetheless, this generated product is the closest from the reference implementation proposed in each reference paper.

Regarding the fourth step, the operating cost will be measured in gas, a unit that represents the cost of performing an atomic operation on an EVM-compatible\footnote{The EVM (Ethereum Virtual Machine) is used by nodes to execute smart contracts.} blockchain.
It is computed by summing all the low-level operations performed during the operation (so-called opcodes).
As the templates of the SPL have been written using Solidity, this metric is very relevant to assessing and comparing the performance of blockchain applications.
However, other metrics might be considered for other technologies.
This aspect is discussed in Subsection \ref{rq-discussion}.

The evaluation of the proposed SPL will be considered satisfying if the products generated from the web platform sufficiently match the requirements formulated by the authors of reproduced applications, and if the gas cost for the deployment and the execution of the generated smart contracts is satisfactory compared to the reference papers implementations.
For the latter, the gas cost of each implementation is shown in Figure \ref{gascost}.

\subsection{Spare Part Study Comparison}

\begin{table*}[t]
  \caption{Spare parts study functional requirements (SR: satisfied in reference paper, SP: satisfied in the generated product).}
  \label{paper1req}
  \begin{tabular}{|c|c|p{12.35cm}|c|c|}
  \hline
  \multicolumn{1}{|c|}{\textbf{Category}} & \textbf{ID} & \textbf{Requirement} & \textbf{SR} & \textbf{SP} \\ \hline
  \multirow{3}{*}{\makecell{Purchase\\request}} & R.1.1 & The engineer shall be able to submit a purchase request. & Yes & Yes \\ \cline{2-5}
  & R.1.2 & The line manager shall be able to approve a purchase request. & Yes & Yes \\ \cline{2-5}
  & R.1.3 & The procurement manager shall be able to approve a purchase request if the requested spare part within the request is missing from the inventory. & Yes & Yes \\ \hline
  \multirow{3}{*}{\makecell{Purchase\\quotation}} & R.1.4 & The procurement manager shall be able to submit purchase quotations for a requested spare part. & Yes & Yes \\ \cline{2-5}
  & R.1.5 & The engineer shall be able to select a purchase quotation for a requested spare part. & Yes & Yes \\ \cline{2-5}
  & R.1.6 & The procurement manager shall be able to confirm the availability of the requested spare part. & Yes & Partially \\ \hline
  \multirow{5}{*}{\makecell{Purchase\\order}} & R.1.7 & The engineer shall be able to submit a purchase order for a requested spare part. & Yes & Yes \\ \cline{2-5}
  & R.1.8 & The line manager shall be able to approve a purchase order. & Yes & Yes \\ \cline{2-5}
  & R.1.9 & The purchase manager shall be able to purchase the spare part specified by the purchase order. & Yes & Yes \\ \cline{2-5}
  & R.1.10 & The engineer shall be able to request a spare part from the inventory. & Yes & Yes \\ \cline{2-5}
  & R.1.11 & The engineer shall be able to submit a purchase order for a requested spare part. & Yes & Yes \\ \hline
  \multirow{2}{*}{\makecell{Spare part\\transfer}} & R.1.12 & An OEM (Original Equipment Manufacturer) shall be able to create a spare part entry. & No & Yes \\ \cline{2-5}
  & R.1.13 & Any participant shall be able to transfer the spare part ownership to another. & No & Yes \\ \hline
  \end{tabular}
\end{table*}

The first study chosen for the evaluation discusses a blockchain-based traceability system for spare parts purchasing in manufacturing \cite{hasan2020blockchain}.
The main motivation for this study is the lack of reliable tracing and tracking of spare parts and their ownership, especially when they are employed in sensible domains, such as aeronautics.
From this study, a set of 13 functional requirements have been identified and classified (Table \ref{paper1req}).
Then, a configuration has been created based on these requirements, and the corresponding product has been generated and deployed to assess its performance.

\subsubsection{Feature Selection}

Two traceability features have been selected. 
The first feature is \texttt{AssetTracking}, as a representation of a spare part must be created by an OEM (Original Equipment Manufacturer) for ownership traceability purposes.
As there is no need for modeling tokenized assets, the \texttt{StructuredAssets} subfeature is used.
Then, the second chosen feature is \texttt{StateMachine}, as it is required to trace the current state of purchasing new spare parts.
Regarding the \texttt{Participants} feature, only individuals have been included in the configuration.
Indeed, there is no need to create groups of individuals (e.g., roles) in this scenario.
The configuration does not include individual types either, as there are no oracle or external services specified.

For storage concerns, the spare parts study does not specify any off-chain storage.
However, events are emitted along the process of refilling spare parts.
Thus, the \texttt{EventsEmission} feature has been included.
Also, the \texttt{StateData} and the \texttt{AssetsData} storage type subfeatures have also been included, due to the specified constraints between features.

\subsubsection{Requirements Satisfaction}

After the generation of a product based on this configuration, the satisfaction of requirements can then be assessed (Table \ref{paper1req}).
In total, 12 of the 13 specific requirements are marked as satisfied.
Indeed, the generated product is able to support these requirements by leveraging a state machine to track the state of spare parts refilling, and the ownership of spare parts through assets.
However, one requirement has been marked as partially filled.
The requirement R6 is difficult to satisfy with the current implementation of the product, as it requires establishing a communication system between the OEM and the procurement manager to ask for spare part availability.
As we only evaluate the on-chain part of both applications (i.e., smart contracts), R.1.4 and R.1.12 have been marked as satisfied.
These requirements demand to store some documents on IPFS (Inter-Planetary File System), a decentralized storage system \cite{benet2014ipfs}, then store the document reference (so-called tag) in the smart contract.
Both the spare-part study implementation and the generated product can do that, however, they do not propose a frontend feature to store a document on IPFS for the moment.
Note that requirements R.1.12 and R.1.13 have been marked as unsatisfied in the spare-part study implementation.
Indeed, only one hardcoded spare part has been found in the spare-part study implementation code, and no function allows the transfer of a spare part from one participant to another.

\subsubsection{Performance Assessment}

\begin{figure}[t]
  \centering
  \includegraphics[width=\linewidth]{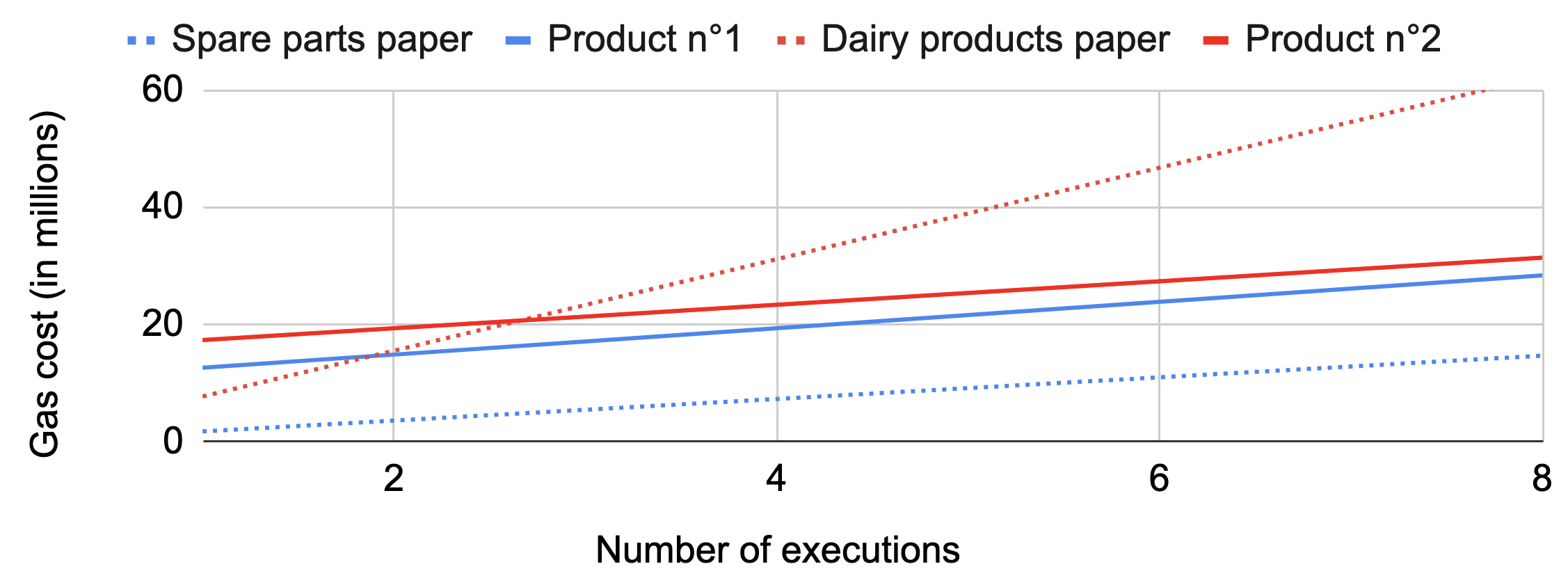}
  \caption{Gas cost of executing several times the reference implementations and generated products.}
  \label{gascost}
\end{figure}

To evaluate the performance ratio between the smart contract proposed in the spare-part study and the generated product, we designed a test scenario for spare part refilling, from the request to the purchase.
This scenario covers the functional requirements specified in Table \ref{paper1req}.
Figure \ref{gascost} compiles the differences from 1 to 8 executions.


The process to compute these metrics is the following.
At first, the cost of deploying the smart contracts is assessed.
This cost is separated from others as usually it is paid only once by the user, at deployment.
However, this is not the case for the spare-part study architecture.
Then, each function was executed, both in the smart contract proposed in the spare-part study and the generated product.
For the latter, the followed scenario involved the creation of an on-chain state machine using the same states as the spare-part study, then transitioning from one state to another providing the same parameters as the first spare part study.

The cost of deploying the generated product is up to 10 431 963 gas, whereas the smart contract proposed in the spare-part study costs 1 513 078 gas to be deployed.
However, the generated product allows the creation of a new traceability process using already deployed contracts, where the smart contract proposed in the spare-part study has to be redeployed to be used when starting a new traceability process.
Thus, the deployment of smart contracts is not a one-time cost in the spare-part study and has to be paid for each traceability process created.
Also, the implementation cost of the generated product includes features for asset management, specified in the spare-part study.
However, these features are missing from the spare-part study implementation.
Regarding the cost of executing the scenario once the deployment is performed, the spare-part study cumulates a gas cost of 329 840, where the generated product adds up to 2 248 064 gas.
Note that two features specified in the requirements are missing from the spare-part study implementation, thus the cost of the generated product for the first 11 requirements can be adjusted to 1 970 268 gas.

Figure \ref{gascost} displays a tendency of the execution cost of both spare-part study implementation and generated products.
To extend these results, we also computed the cost of executing the scenario several times.
As for the spare-part study implementation, the cost is obtained for $N$ execution by summing $N$ time for both the deployment and the function execution cost.
In the product generated, the cost is obtained by summing $N$ times the function execution cost and only then adding the deployment cost.
Even if these cost models appear simplistic at first sight, they correctly represent the actual cost for EVM-based languages such as Solidity, since EVM imposes a deterministic execution model and fixed code deployment costs.
This difference in calculation method is explained as the generated product's smart contracts do not have to be redeployed to create a new process.
More rationale on identified cost differences between these two implementations is given in Section \ref{discussion}.

\subsection{Dairy Products Study Comparison}

For the second chosen study, a blockchain-based food supply chain traceability for dairy products is introduced \cite{casino2021blockchain}. 
As safety is a critical aspect of food supply chains, blockchain and smart contracts can be used to build a secure and trustworthy architecture for food supply chain traceability.
In their work, Casino et al. propose such architecture through a concrete use case for the traceability of dairy products.
Eleven functional requirements have been identified and classified in this study (Table \ref{paper2req}).
From these requirements, a configuration has been made on the web platform, and the corresponding product has been generated for comparison.

\subsubsection{Feature Selection}

This case study both involves the tracking of asset ownership, records, and state changes in a process.
The three different tracking methods have been selected to address these requirements: \texttt{AssetTracking}, \texttt{RecordsHistory}, and \texttt{StateMachine}.
Also, as there is no need for modeling tokenized assets, the \texttt{StructuredAssets} subfeature is used.

For the \texttt{Participants} main feature, the paper describes the need to define two roles.
The first one is the \textit{Stakeholders} role: stakeholders are involved in the milk transformation process.
The second one is the \textit{Administrators} role.
Members of this role group are employees from the dairy company and oversee the blockchain traceability application.
They are able to perform administration operations, such as adding new stakeholders.
The presence of roles in this application justifies the selection of the \texttt{Role} feature.
According to the dairy products study, it is also possible to create new stakeholders or delete them at any moment.
This involves the following features and their subfeatures: \texttt{CreateIndividual}, \texttt{DeleteIndividual}, and \texttt{AddRole}.
However, \texttt{IndividualTypes} have not been added into the configuration, as there is no explicitly mentioned oracles or services.

Regarding the \texttt{Storage} feature, the study does not mention the emission of events.
As this is an expensive feature in terms of gas cost, \texttt{EventEmission} has been excluded from the configuration.
The other storage subfeatures, notably the ones related to the traceability data, were automatically selected.

\subsubsection{Requirements Satisfaction}

Once the configuration step is finished, we assessed the satisfaction of the extracted requirements (Table \ref{paper2req}).
In total, 8 out of 11 requirements have been marked as satisfied, 2 requirements marked as partially satisfied, and one requirement marked as unsatisfied.
In our SPL, an asset can only be weakly attached to a process (here, state machine instances), using the additional data field to reference the instance.
Thus, the requirements R.2.8 and R.2.9 have also both been marked as partially satisfied.
Regarding the requirement R.2.7, it is not satisfied as it demands a feature to stale an ongoing process: this aspect is not handled by the generated product.
Also, as we only evaluate smart contracts in the evaluation, the requirement R.2.5 is satisfied.
Indeed, as in the first study, it is possible to attach an IPFS tag to an asset in the generated product, to link it to a document.
However, this requires uploading the document beforehand, a feature not handled by the web platform at the moment.

\begin{table*}[t]
  \caption{Dairy products study functional requirements (SR: satisfied in reference paper, SP: satisfied in generated product).}
  \label{paper2req}
  \begin{tabular}{|c|c|p{11.7cm}|c|c|}
  \hline
  \multicolumn{1}{|c|}{\textbf{Category}} & \textbf{ID} & \textbf{Requirement} & \textbf{SR} & \textbf{SP} \\ \hline
  \multirow{5}{*}{\makecell{Product\\management}} & R.2.1 & An administrator or a stakeholder shall be able to create a product. & Yes & Yes \\ \cline{2-5}
  & R.2.2 & An administrator shall be able to change the stakeholder of a product. & Yes & Yes \\ \cline{2-5}
  & R.2.3 & A stakeholder shall be able to change the stakeholder of a product if owned. & Yes & Yes \\ \cline{2-5}
  & R.2.4 & An administrator or a stakeholder shall be able to push a new record for a given product. & Yes & Yes \\ \cline{2-5}
  & R.2.5 & An administrator or a stakeholder shall be able to attach a chemical test to a product. & Yes & Yes \\ \hline
  \multirow{4}{*}{\makecell{Milk\\transformation\\process\\management}} & R.2.6 & An administrator or a stakeholder shall be able to create a new milk transformation process. & Yes & Yes \\ \cline{2-5}
  & R.2.7 & An administrator or a stakeholder shall be able to disable a milk transformation process. & Yes & No \\ \cline{2-5}
  & R.2.8 & An administrator shall be able to link a product to a milk transformation process. & Yes & Partially \\ \cline{2-5}
  & R.2.9 & A stakeholder shall be able to link a product to a milk transformation process if owned. & Yes & Partially \\ \hline
  \multirow{2}{*}{\makecell{Stakeholder\\management}} & R.2.10 & An administrator shall be able to disable a stakeholder. & Yes & Yes \\ \cline{2-5}
  & R.2.11 & An administrator shall be able to create a new stakeholder. & Yes & Yes \\ \hline
  \end{tabular}
\end{table*}

\subsubsection{Performance Assessment}

The cost of deploying the smart contracts is assessed, then each function was executed, both in the smart contract proposed in the first reference paper and the generated product.
Figure \ref{gascost} shows the resulting costs from 1 to 8 executions.
The cost of deploying the generated product is up to 15 400 174 gas, whereas the smart contract proposed in the dairy products study costs 6 748 484 gas to be deployed.
As in the first paper, this is not a one-time cost for the dairy products study implementation: smart contracts have to be redeployed for each legal agreement signed between stakeholders and the dairy company in charge of the application.
For the functions-related costs, the dairy products study implementation sums up a gas cost of 1 044 928, and the generated product a gas cost of 2 004 322.
As in the spare parts study comparison, we also computed the cost of executing the scenario several times.
Like the first paper, the implementation cost is added only once to the generated product total gas cost, whereas it is added $N$ times for the dairy products study implementation.


\section{Discussion}
\label{discussion}

\subsection{Research Questions}
\label{rq-discussion}

In the evaluation section, the relevance of the approach is assessed by replicating on-chain traceability applications found in other works using the web platform.
The gas cost of generated products was also assessed by comparing it to the gas cost found for the reference studies implementations.
This section discusses these results in the light of formulated research questions (Section \ref{research-questions}).

\subsubsection{RQ1}

To address the first research question, we compared the requirements satisfaction rate of the generated products and the reference papers code.
Indeed, if it is possible to replicate most of the existing blockchain-based traceability applications by only using the web platform, the SPL approach is relevant.
It has been shown that the web platform was able to produce blockchain applications that satisfy most of the requirements expressed by the studies that were used as reference.
Yet, some of the requirements were not fully satisfied.
A reason is the genericity of the products that can be generated by the web platform.
Indeed, the templates have been designed to be flexible rather than implementing specific domain-oriented features.
An illustration of this flexibility is the management of roles: rather than using data structures tailored after the possible roles in a traceability application, a generic data structure named Role is implemented.
Also, domain-specific features might be missing from the generated product.
This has been faced during the evaluation (Section \ref{evaluation}), where some requirements need to verify a specific condition or execute a defined operation before changing the traceability process state.
Nevertheless, the design of the SPL facilitates the integration of new domain-oriented features.
In this case, the generated product is solid ground to start implementing more complex features on it.

\subsubsection{RQ2}

The second research question consists of evaluating if differences between applications generated from an SPL or implemented using a traditional software engineering approach exist.
For these applications, the gas cost of deploying and then executing several times a defined scenario has been measured.
Then, the divergence of design and code between these applications has been studied to explain the measured gas costs.
In the spare parts study, the generated product was more expensive to deploy and execute several times, and for the dairy products study, less expensive after 3 executions.
This difference is mainly due to two architectural aspects: the redeployment of smart contracts when willing to relaunch a new traceability process, and the deployment of numerous contracts to facilitate contract upgradeability.
Indeed, redeploying a contract requires reallocating a large amount of storage to initialize state variables and store the source code.
The products generated by the web platform are designed to avoid this issue: a new state machine (by extension, a traceability process) can be created using a dedicated function.
The separation of concerns between data and logic also addresses this issue, as a new controller can be deployed to upgrade some features in generated products rather than redeploying everything.
However, this approach has a drawback: the logic required for the separation of concerns and easier upgradeability requires the deployment of larger smart contracts.
This results in a more expensive deployment for generated products.

The implementation of the generated products features and reference study implementations also differs.
For the latter, hardcoded values were found, notably for the definition of participants and roles.
This leads to decreased gas costs, as there is no additional feature for participants dynamic management.
On the opposite, the generated products derived from our SPL are flexible and foster maintainability and upgradeability.
The flexibility of this approach increases the operating costs of the application.
Nevertheless, the high gas costs observed during the evaluation of the SPL might be reduced in future works by implementing features to reduce the code and needed storage size, to the detriment of upgradeability.
Also, the deployment of smart contracts and the execution of functions is free on private blockchains networks, such as Proof-of-Authority-based Ethereum networks.
In this context, it is not always necessary to optimize the application for gas-cost reduction.

It should also be noted that although the gas cost is an accurate metric to describe Ethereum-based smart contract performance, it is not systematically generalizable to any blockchain technology.
Indeed, there is no gas cost at all on other non-EVM-based blockchains, such as Hyperledger Fabric.
Other metrics might be considered to assess the performance of blockchain applications in future works using these technologies.
For instance, the resource usage of an application (e.g., CPU, RAM, storage size, etc.) could be monitored.
The cost of executing the features themselves could also vary depending on the blockchain used.
As an example, a feature for data confidentiality requires to implement a function to encrypt data on Ethereum-based blockchains.
On Hyperledger Fabric, this is unnecessary as it is possible to restrict the read access of a contract to a defined set of participants, using channels \cite{androulaki2018hyperledger}.

\subsection{Lessons Learned}
\label{lessons-learned}

The main advantage identified during the completion of the study was the time saved compared to manually developing traceability applications.
Indeed, after the identification of desired requirements in these works, the configuration and the generation of blockchain applications can be done in minutes.
Also, the quality of generated products benefits from the integration of good practices, design patterns, and standards in core assets.
However, the main drawback to this approach is the time overhead needed to set up the SPL (feature analysis, FM development, templates development).
Compared to a traditional software engineering approach, additional time has been needed to extract the features from the literature, structure them in a feature model, and implement the templates taking in mind the possible combinations between the features as well as the organization between smart contracts (e.g. separating logic and data).
Furthermore, this work does not address the challenges of SPL evolution, as the web platform was built over a new feature model.
Additional tasks not carried out in this work might be needed to handle the SPL evolution, as illustrated in \cite{abbas2020product}.
This approach might not be tailored for a company willing to implement a particular blockchain solution, but might suit IT services companies that want to provide a wide range of blockchain products with shared commonalities to their customers.

The templating engine used in this contribution was enough to illustrate the capability of generating blockchain products from configurations.
However, a domain engineer may feel limited by the templating engine when implementing many templates for large-scale SPLs.
The implementation of the different features within templates might also be tedious, as it must take into account all the possible combinations of features and possible nestings.

Nonetheless, this issue can be mitigated in the blockchain field by different means.
First, smart contracts can be designed in a way that the resulting architecture is a set of loosely coupled smart contracts.
This approach eases the addition of new features to the SPL.
Such architecture is notably introduced by Tonelli et al. \cite{tonelli2019implementing}, as they implement a microservices system with blockchain smart contracts.
Consequently, the smart contract architecture proposed in this work was designed with modularity as a main concern.
Second, many design patterns, standards, and commonly reused code blocks already exist.
As identified by Chen et al., 26\% of Ethereum smart contracts code block are from reused sources, notably, ERC20-related contracts \cite{chen2021understanding}.
Indeed, ERC20\footnote{\url{https://ethereum.org/en/developers/docs/standards/tokens/erc-20}} is a standard for the creation of fungible tokens on Ethereum.
This existing code can be easily bundled into a feature, reusable in many SPLs.

\subsection{Research Challenges}

Using SPLs to create blockchain applications raises new research challenges to address.
In this paper, the Solidity language has been chosen to develop smart contracts.
However, a wider range of languages exists to develop smart contracts for one or other blockchain technologies (e.g., Solidity, Rust, etc.).
Future FMs of blockchain products could contain a feature for the selection of a specific smart contract language.
This feature could yield SPLs that are able to produce the same application for multiple blockchain technologies.
It would allow developers to focus on the application to build rather than the blockchain target behind and its technical specificities.
Still, there is an issue with the implementation of such features: the programming model might differ between different blockchains.
For instance, Ethereum is account-based, whereas other blockchains such as Bitcoin, rely on a UXTO model \cite{brunjes2020utxo}.
A consequence of these different paradigms could be the impossibility to design some features with specific blockchain technologies.
Therefore, more research is needed to assess the generalisability of this approach to other domains or other SPL paradigms.

Also, this paper proposes a domain-oriented FM (on-chain traceability), yet another type of FM could be created around existing blockchain features.
The resulting SPL could allow the creation of generic blockchain applications that provide a solid ground for developers to start implementing domain features above.
The evolution of SPLs, when core assets (e.g., templates, FMs) evolve over time to address newer requirements or changes in the technology used \cite{marques2019software}, is also a challenge for blockchain SPLs.
This issue is very relevant to blockchain: due to the novelty of the field, many existing standards, patterns, and commonly reused block codes might change in the future, impacting existing features.
As mentioned in Subsection \ref{lessons-learned}, this is an issue that companies might face when experimenting SPLs for blockchain applications.
Future research on blockchain-based SPLs should consider this issue and include mechanisms to handle the evolution of blockchain core assets.

\section{Conclusion}
\label{conclusion}

As the development of blockchain applications is still tedious and error-prone, the usage of an SPL can help in the systematic reuse of existing code, good practices, and standards (e.g., Ethereum ERCs) to build robust and efficient applications.
This paper denotes the relevance of leveraging SPLs for the design and implementation of blockchain-based applications with an exemplified approach.
First, a feature model for on-chain traceability applications is introduced, built by extracting features from 5 different works in this field.
From this model, a web platform is proposed to allow the configuration of an on-chain application.
The web platform also includes a code generator that reuses this configuration to feed a templating engine that produces a working blockchain application, without any coding. By specifying its desired features, the user is capable of generating an application for on-chain traceability that suits its needs.
Also, the produced code is designed to be highly modular, thus easing the addition of new features, either through adding extra features in the feature model or manually.
This approach is validated by using the web platform to recreate existing on-chain traceability applications proposed in the literature.

Many research challenges still have to be addressed, such as the management of the SPL evolution considering the rapid pace of blockchain development.
Yet, this paper paves the way for blockchain-backed solutions created with the SPL method.

\section*{Open Science}

Following open science principles, the source code associated to web platform introduced in this paper and the results obtained in the evaluation of the contribution are available on GitHub\footnote{\url{https://github.com/harmonica-project/BANCO}}. A rationale is also given to change the proposed feature model and its related code templates to other domains and purposes.

\section*{Acknowledgements}

This work is partially supported by the Natural Sciences and Engineering Research Council of Canada (NSERC) grant RGPIN-2017-05421, and the Mitacs Globalink Research Award grant IT30564.

\balance
\bibliographystyle{ACM-Reference-Format}
\bibliography{v2bibliography.bib}

\end{document}